\newcommand{\tw}{t_{\rm w}}
\newcommand{\Tc}{T_{\rm c}}
\newcommand{\Tm}{T_{\rm m}}
\newcommand{\Vc}{V_{\rm cluster}}

\documentclass[aps,prb,twocolumn]{revtex4}

\usepackage{graphicx}% Include figure files

\def\nle{\ \raise.3ex\hbox{$<$}\kern-0.8em\lower.7ex\hbox{$\sim$}\ }
\def\nge{\ \raise.3ex\hbox{$>$}\kern-0.8em\lower.7ex\hbox{$\sim$}\ }

\begin{document}

\title{Aging and Memory Effects in Superparamagnets and Superspin Glasses
}

\author{M. Sasaki}
\affiliation{Institute for Solid State Physics, University of Tokyo, 
5-1-5 Kashiwa-no-ha, Kashiwa, Chiba 277-8581, Japan}

\author{P. E. J{\"o}nsson}
\affiliation{Institute for Solid State Physics, University of Tokyo, 
5-1-5 Kashiwa-no-ha, Kashiwa, Chiba 277-8581, Japan}

\author{H. Takayama}
\affiliation{Institute for Solid State Physics, University of Tokyo, 
5-1-5 Kashiwa-no-ha, Kashiwa, Chiba 277-8581, Japan}

\author{H. Mamiya}
\affiliation
{National Institute for Materials Science, Sengen 1-2-1, Tsukuba, Ibaraki 305-0047, Japan
}

\date{\today}
 
\begin{abstract}
Many dense magnetic nanoparticle systems exhibit slow
 dynamics which is qualitatively indistinguishable from that observed in
 atomic spin glasses and its origin is attributed to 
dipole interactions among particle moments (or superspins). 
However, even in dilute nanoparticle systems where the  dipole
 interactions are vanishingly small, slow dynamics is
 observed and is attributed solely to a broad 
 distribution of relaxation times which in turn comes from that of the
 anisotropy energy barriers. To  clarify characteristic differences
 between the two types of slow dynamics, we study a simple model of
 a non-interacting  nanoparticle system (a superparamagnet) analytically
as well as ferritin (a superparamagnet) and a dense 
Fe-N nanoparticle system (a superspin glass) experimentally. 
It is found that superparamagnets in fact show aging (a waiting time
 dependence) of the thermoremanent-magnetization as well as various
 memory effects. We also find some dynamical phenomena peculiar only
 to superspin glasses such as the flatness of the field-cooled magnetization 
below the critical temperature and memory effects in the zero-field-cooled 
magnetization. These dynamical phenomena are qualitatively reproduced 
by the random energy model, and are well interpreted by the so-called 
droplet theory in the field of the spin-glass study. 
\end{abstract}

\maketitle
 
\section{Introduction}
One of the most attractive topics in the field of condensed matter physics 
is slow dynamics such as non-exponential relaxation, aging 
(a waiting time dependence of observables),\cite{Struik78,Lundgren83} 
and memory effects. These phenomena are observed in various systems like 
polymers,\cite{Struik78,BellonCiliberto99,BellonCiliberto00}
high-$T_{\rm c}$ super-conductors,\cite{RosselMaeno89}
granular materials\cite{Josserand00} and spin glasses. 
Especially, in the field of spin glasses, slow dynamics has been studied 
widely both experimentally\cite{Lundgren83,VincentBouchaud95,NordbladSvedlindh98,JonasonVincent98} and 
theoretically\cite{CugliandoloKurchan93,BouchaudCugliandolo98,YoshinoLemaitre00,SasakiNemoto00,KomoriYoshino00,BerthierBouchaud02} to examine the validity 
of novel concepts of spin glasses such as a hierarchical organization of 
states\cite{MezardParisi84,MezardParisi84b} and 
temperature chaos.\cite{FisherHuse86,BrayMoore87,FisherHuse88} 
These extensive studies have revealed various interesting behavior in dynamics 
like coexistence of memory and 
rejuvenation.\cite{VincentBouchaud95,NordbladSvedlindh98,JonasonVincent98} 
Such findings have stimulated many researchers to study 
slow dynamics in various systems 
like geometrically frustrated magnets,\cite{WillsDupuis00,DupuisVincent02b} 
transition-metal oxides,\cite{MathieuSvedlindh00} 
orientational glasses,\cite{AlberichDoussineau97,AlberichBouchaud98} 
supercooled liquids,\cite{LehenyNagel98} and dense magnetic nanoparticle systems\cite{JonssonMattsson95,MamiyaNakatani99,JonssonHansen00,Sahoo02,Sahoo03,Jonsson04}
by using experimental protocols developed in the study of spin glasses. 
Magnetic nanoparticle systems, which we study in this paper, 
are of current interest because of their significance for technological applications as well as for their fundamental magnetic properties.\cite{BatlleLabarta2002}

In magnetic nanoparticle systems, there are two possible origins of slow 
dynamics. The first one is a broad distribution of relaxation times 
originating solely from that of the anisotropy energy barriers of 
each nanoparticle moment. This is the only source of slow dynamics 
for sparse (weakly interacting) magnetic nanoparticle systems, in which the nanoparticles are fixed in space. 
We hereafter call magnetic moments of each nanoparticle {\it superspins}, 
and such weakly interacting magnetic nanoparticle systems
{\it superparamagnets}. 
However, for dense magnetic nanoparticle systems, 
there is a second possible origin of slow dynamics, 
namely, cooperative spin glass dynamics due to frustration 
caused by strong dipolar interactions among the particles and randomness in 
the particle positions and anisotropy axes orientations.\cite{LuoNagel91} 
In fact, evidences for a spin glass transition such as 
the critical divergence of the nonlinear susceptibility have been 
found in dense magnetic nanoparticle 
systems.\cite{JonssonSvedlindh98,MamiyaNakatani99B,Sahoo02b} 
We hereafter call such dense magnetic nanoparticle systems, 
which exhibit spin glass behavior, {\it superspin glasses}.

Now the point is that magnetic nanoparticle systems involve 
two possible mechanisms for slow dynamics, and which of the two is relevant 
depends essentially on the concentration of nanoparticles. 
Then, in order to understand appropriately slow dynamics 
in magnetic nanoparticle systems, 
it is desirable to clarify which observed phenomena are simply due 
to slow dynamics caused by a broad distribution of relaxation times, 
and which ones are brought by cooperative dynamics peculiar to superspin 
glasses. For this purpose, we first study a simple model of non-interacting 
magnetic nanoparticle systems (superparamagnets) analytically. 
As a consequence, we find that even superparamagnets exhibit 
aging of the thermoremanent-magnetization and various memory effects. 
In particular, we show that the curious memory effects recently reported 
by Sun et al.,\cite{Sun03} which were claimed to give evidences 
of the existence of a superspin glass phase, 
can be understood simply as superparamagnetic relaxation (see also 
Refs.~\onlinecite{SasakiJonsson03,ZhengZhang04,ChakravartyFrydman04}.)

We also perform experiments on a ferritin 
(a superparamagnet\cite{KilcoyneCywinski95,MamiyaNakatani02}) 
and a dense Fe-N nanoparticle systems (a superspin glass\cite{MamiyaNakatani98,MamiyaNakatani99,MamiyaNakatani99B,JonssonYoshino04}). 
The results of ferritin are qualitatively similar to those of 
our simple model of superparamagnets. 
The comparison of the phenomena observed in the superparamagnet and 
the superspin glass reveals some properties peculiar only to 
superspin glasses, e.g., the flatness of the field-cooled magnetization
below the critical temperature and memory effects in the
zero-field-cooled magnetization. Particularly, the former phenomenon
reminds us of Parisi's equilibrium susceptibility in the spin-glass
mean-field theory.\cite{Parisi83} However, we propose an interpretation
based on the spin-glass droplet theory\cite{FisherHuse88B,FisherHuse88}
which predicts the instability of the spin-glass phase under a static
magnetic field of any strengths and so claims the observed field-cooled
magnetization to be a property 
far from equilibrium.\cite{TakayamaHukushima03} 
We also show that 
these experimental results peculiar to superspin glasses are
qualitatively reproduced by the random energy 
model.\cite{Derrida81,Bouchaud92,BouchaudDean95}

The outline of the present manuscript is as follows. In 
section~\ref{sec:2} we introduce a model of superparamagnets and
report aging and memory effects observed 
in this model. The results of experiments on ferritin 
are also shown in this section. 
In section~\ref{sec:3} we show experimental results on a dense Fe-N 
nanoparticle system. Some properties found only in the superspin glass 
are interpreted by the random energy model and the droplet theory. 
Section~\ref{sec:4} is devoted for summary. 

\section{Slow Dynamics in Superparamagnets}
\label{sec:2}
\subsection{Model and master equation approach}
\label{subsec:2-1}
Here we adopt a simple model which is considered to describe the essential 
slow dynamics in non-interacting 
magnetic nanoparticle systems (superparamagnets). The 
magnetic moment (superspin) of one nanoparticle, which does not interact 
with any other superspins, is supposed to occupy one of two states with 
energies $-K V\pm h M_{\rm s} V$, where $K$ is the bulk anisotropy constant, 
$V$ the volume of the nanoparticle, $h$ an applied field in linear 
response regime, and $M_{\rm s}$ the saturation magnetization. 
Here we supposed that the direction of the field is parallel to 
the anisotropy axes for simplicity. 
The superparamagnetic relaxation time in zero field 
for the thermal activation over the energy barrier $KV$ is given by 
$\tau=\tau_0\exp(KV/T)$, where $\tau_0$ is a microscopic time. 

The occupation probability of one of the two states, in which the superspin 
is in parallel (antiparallel) to the field direction, is denoted by 
$p_1(t)\ (1-p_1(t))$, and is solved by the following 
master equation\cite{KlikChang94}
\begin{equation}
{d \over dt}p_1(t) = - W_{1\rightarrow2}(t) p_1(t)
+W_{2\rightarrow1}(t) \{1-p_1(t)\},
\label{eqn:master}
\end{equation}
where $W_{1\rightarrow2}(t)\ (W_{2\rightarrow1}(t))$ is the 
transition rate from the state 1 to 2 (2 to 1) at time $t$. To the
leading order in $h(t)$ they are written as
\begin{eqnarray}
W_{1\rightarrow2}(t) = \frac12 \tau_0^{-1} {\rm exp}[-KV/T(t)]\{1-M_{\rm s}Vh(t)/T(t)\}, 
\label{eqn:TransitionRateA}
\\
W_{2\rightarrow1}(t) = \frac12 \tau_0^{-1} {\rm exp}[-KV/T(t)]\{1+M_{\rm s}Vh(t)/T(t)\}.
\label{eqn:TransitionRateB}
\end{eqnarray}
The above master equation can be solved analytically for any 
temperatures and field protocols represented by $T(t)$ and $h(t)$ from 
a given initial condition, and the magnetization of the particle 
with volume $V$ is given by 
\begin{equation}
M(t;V) = [2p_1(t;V)-1]M_{\rm s}V.  
\label{eqn:mag}
\end{equation}
For example, in the case that $h(t)=h$ and $T(t)=T$, we obtain
\begin{eqnarray}
M(t;V)&=&M(t=0;V)\exp(-t/\tau)\nonumber \\
&&+\frac{(M_{\rm s}V)^2h}{T}\{1-\exp(-t/\tau)\},
\label{eqn:mag_const}
\end{eqnarray}
where $\tau\equiv \tau_0\exp(KV/T)$. 
Note that the additional condition $h=0$ leads us to the familiar formulation 
for the decay of the thermoremanent-magnetization. 

From Eqs.(1-4), 
we notice that $p_1(t)=1/2 \ (M(t)=0)$ at any $t$ if $p_1(0)=1/2$ and 
$h(t)=0$. This means that in any {\em genuine} zero-field-cooled 
(ZFC) processes starting from $M=0$, $p_1(t)$ is independent of
the schedule of temperature change $T(t)$, i.e., no memory is imprinted 
in the process. 
Experimentally, a demagnetized initial state is obtained by choosing 
the starting temperature sufficiently high.

The total magnetization of the nanoparticle system is evaluated by
averaging over the volume distribution,
\begin{equation}
{\overline M(t)} = \int dV P(V)M(t;V) \equiv \int dV M_{\rm spec}(t;V).
\label{eqn:ave-mag}
\end{equation}
Here, the integrand (the $M$-spectrum) denoted as $M_{\rm spec}(t;V)$
plays an important role in the arguments below.
For the explicit 
evaluation of ${\overline M(t)}$, we use a log-normal distribution   
\begin{equation}
P(V)=\exp[-\ln(V)^2/(2\gamma^2)]/(\gamma V \sqrt{2 \pi}), 
\label{eqn:distrib}
\end{equation}
with $\gamma=0.6$. Although quantitative and some minute qualitative
results may depend on the value of $\gamma$, the functional form of
$P(V)$, and even our basic assumption of the two-states representation,
we do not go into such details here, expecting that our simplest model
catches up the essence of slow dynamics of superparamagnets. 

In the present work the average anisotropic energy $K{\overline V}$ 
is chosen as the unit of energy as well as that of temperature by setting 
$k_{\rm B}=1$. 
$V$ is measured in unit of the average volume ${\overline V}$, which for the log-normal distribution [Eq.(\ref{eqn:distrib})] is given by $\exp(\gamma^2/2)$. 
As for the time-scale, we suppose that the microscopic 
time $\tau_0$ for superspins of realistic nanoparticles 
is around $10^{-9}$~s, and that a typical experimental time window is 
around 10$^2$~s. We therefore investigate our model in the time window around 
$10^{11}$~$\tau_0$ expecting that it 
corresponds to typical experimental time scales. 

\begin{figure}[h]
\includegraphics[height=\columnwidth,angle=270]{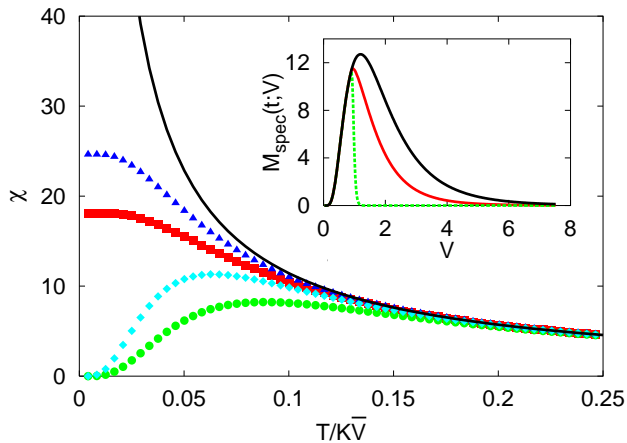}
\caption{(Color online) ZFCM/FCM with the cooling rate $r=2.4\cdot 10^{12}~\tau_0$ 
per temperature unit (circles/squares) and 
those with slower cooling rate $r=2.4\cdot 10^{16}$~$\tau_0$ 
(diamonds/triangles). The line is the susceptibility in equilibrium 
(the Curie law). The inset shows the $M$-spectra of the ZFCM, 
FCM and the magnetization in equilibrium at $T=0.042$ (from left to right).
The cooling rate for ZFCM/FCM is $2.4\cdot 10^{12}~\tau_0$. 
}
\label{fig1}
\end{figure}
\subsection{ZFC and FC magnetizations}
\label{subsec:2-2}
Let us begin our arguments from the most fundamental and well-known 
protocols, i.e., the measuring processes of the zero-field-cooled 
magnetization (ZFCM) and the field-cooled magnetization (FCM). In the ZFC 
process, the system is rapidly cooled to a low temperature 
in zero field, and then the induced magnetization by an applied field 
$h$ is measured as the temperature is gradually increased. 
In the FC process, on the other hand, the system is 
gradually cooled under $h$ from a sufficiently high temperature so that 
the system is in equilibrium at the highest temperature. The circles/squares 
in Fig.~\ref{fig1} represent the ZFCM/FCM observed with heating/cooling rate, 
$r$, of 2.4$\times 10^{12}$~$\tau_0$ per temperature unit.\footnote[1]{
More explicitly, the master equation (\ref{eqn:master}) is solved 
by changing the temperature 
stepwise with a step of $\Delta T=4.2 \times 10^{-3}$ and at each 
temperature we let the system relax for a period of 
$\Delta t = 1.0 \times 10^{10}$~$\tau_0$.}  
As usually adopted, the peak position 
of the ZFCM is regarded as the blocking temperature, $T_{\rm B}$, 
which is $\simeq 0.088$ for the present process. If the rate $r$ is 
$10^4$ times slower, we obtain $T_{\rm B} \simeq 0.063$ (diamonds for 
the ZFCM and triangles for the FCM). 
If we make $r$ infinitely slow, 
both the ZFCM and the FCM curves coincide with the one given by the Curie law. 

In the inset of Fig.~\ref{fig1} we show the $M$-spectra (the integrand of 
Eq.(\ref{eqn:ave-mag})) of the ZFCM, FCM and the magnetization in 
equilibrium at $T=0.042$ (from left to right). One can clearly see that 
the parts of the three $M$-spectra 
for $V$ smaller than a certain value, which we denote as $V_{\rm B}$, lie 
on top of each other. This means that superspins of these small nanoparticles 
are equilibrated within the characteristic time-scale of the cooling/heating 
process.  % or $\delta t$ introduced above. 
On the other hand, the $M$-spectrum 
of the ZFCM at $V \nge V_{\rm B}$ is zero, indicating that superspins of 
these larger nanoparticles are still blocked to their initial values. 
We call $V_{\rm B}$ 
the blocking volume which depends strongly (linearly) on $T$ and 
weakly (logarithmically) on the observation time-scale. 
Also we call superspins of nanoparticles with $V \nle V_{\rm B}$, 
$V \simeq V_{\rm B}$, and $V \nge V_{\rm B}$ {\it superparamagnetic}, 
{\it dynamically active}, and {\it blocked} or {\it frozen}, respectively. 

By passing we emphasize another characteristic feature of the FCM in 
superparamagnets. Namely, the FCM always increases as 
the temperature is decreased. This is simply because superspins are blocked 
(or frozen) in the direction of the field. 

\begin{figure}[b]
\includegraphics[height=\columnwidth,angle=270]{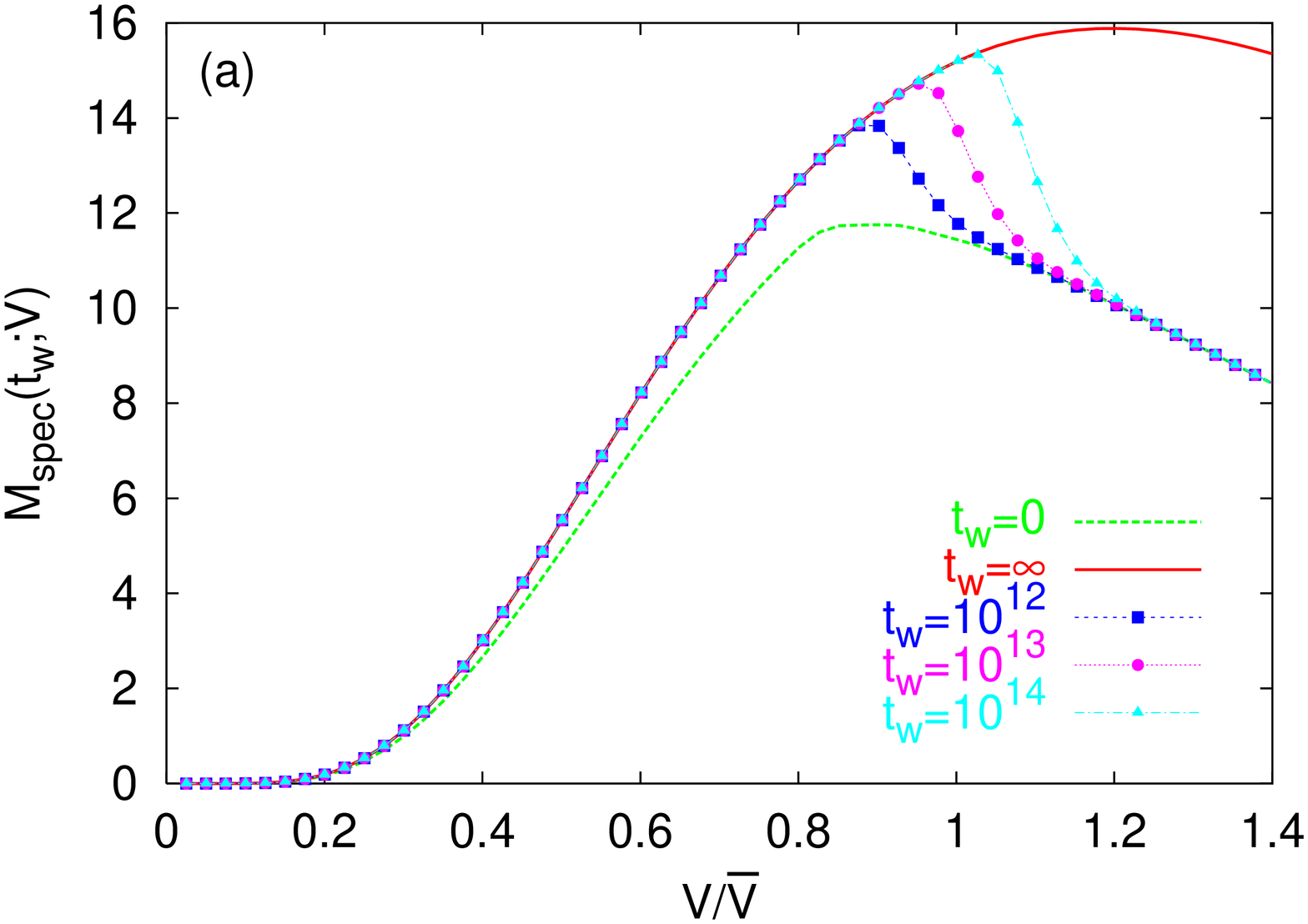}
\includegraphics[height=\columnwidth,angle=270]{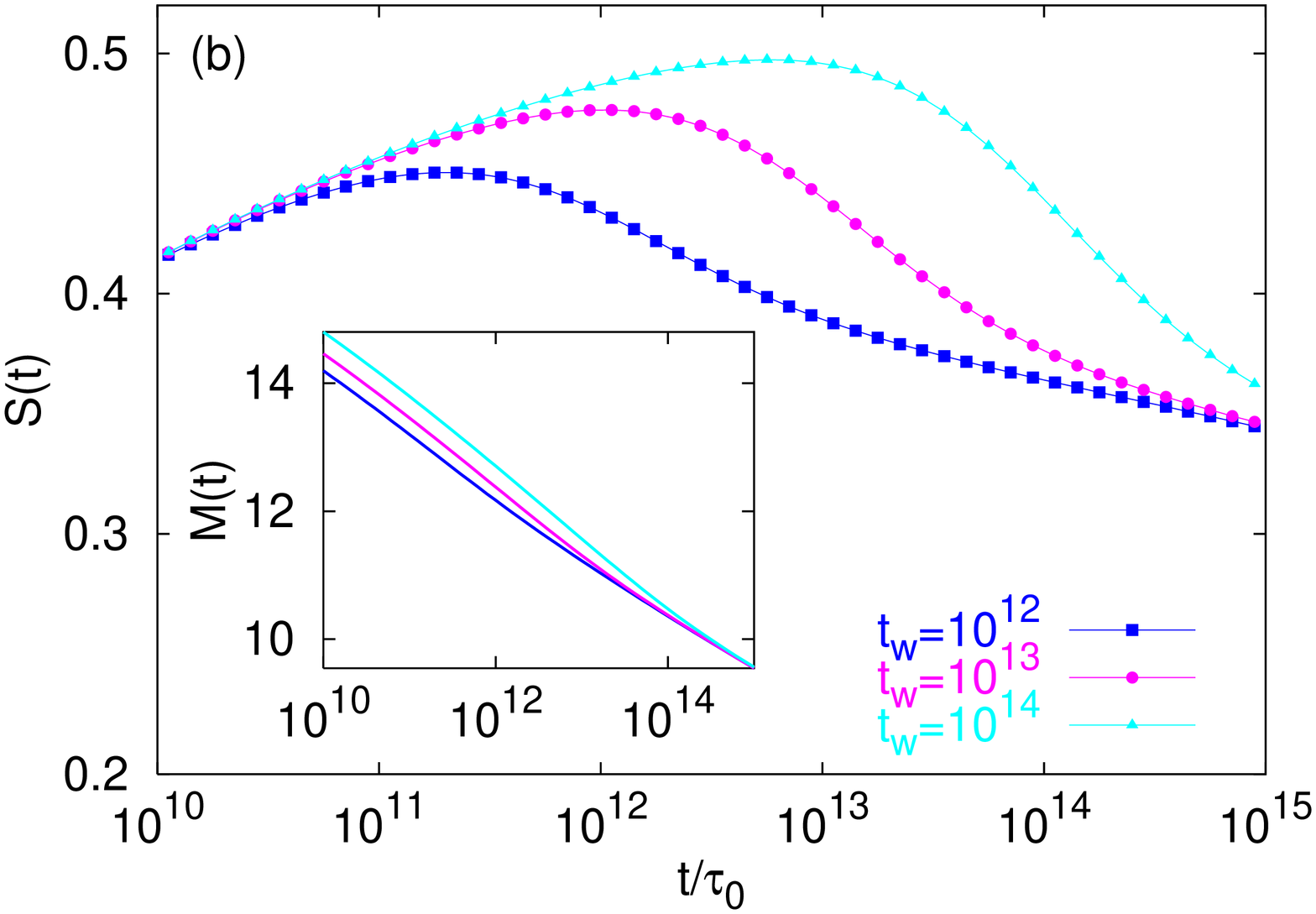}
\caption{(Color online) (a) $M_{\rm spec}(\tw;V)$ of the FC process. 
The system is cooled to $\Tm \ (=0.033$) at the rate of 
$2.4\cdot 10^{12}$~$\tau_0$ per temperature unit, 
and is kept at $\Tm$ for $\tw$. The field is applied in the whole process. 
(b) Susceptibility $\chi_{\rm TRM}(t,t_{\rm w})$ measured in the TRM protocol 
(inset) and its logarithmic time derivative 
$S(t)\equiv -d\log {\chi}_{\rm TRM}(t,t_{\rm w}) /d\log t$ (main frame) 
vs t, where $t$ is the elapsed time after the field is cut. 
The cooling rate and $\Tm$ are the same as in a). In the inset, 
the corresponding waiting time increases from left to right.}
\label{fig2}
\end{figure}

\subsection{Aging and memory effects}
\label{subsec:2-3}
Let us now consider the thermoremanent-magnetization (TRM) protocol, 
where we cool the system in a field $h$ at a certain rate, 
stop the cooling at a measurement temperature, $\Tm$,  let 
the system relax for a waiting time of $\tw$, and then cut the field and 
observe the magnetization decay. 
During the FC aging before cutting the field, 
%In the period before the field is cut, 
the parts of the $M$-spectrum for the frozen and superparamagnetic superspins, 
%introduced in the preceding subsection, 
do not change significantly, 
while that of the dynamically active superspins does change as seen 
in Fig.~\ref{fig2}(a). 
%Now the important point is that 
The peak of the $M$-spectrum shifts 
to larger volumes  with increasing $\tw$. The peak position appears around 
$V^*$ where the corresponding relaxation time $\tau_0\exp(KV^*/\Tm)$ is 
comparable with $\tw$. This naturally means that the TRM decreases most 
rapidly when the time $t$ elapsed after the field is cut is nearly equal to 
$\tw$. Indeed, Fig.~\ref{fig2}(b) shows that the relaxation rate 
$S(t)\equiv -h^{-1}d\log M /d\log t$ in the TRM protocol has a peak 
around $\tw$. Thus we conclude that aging (a $\tw$-dependence) of 
the TRM does exist even in superparamagnets.

\begin{figure}[b]
\includegraphics[height=\columnwidth,angle=270]{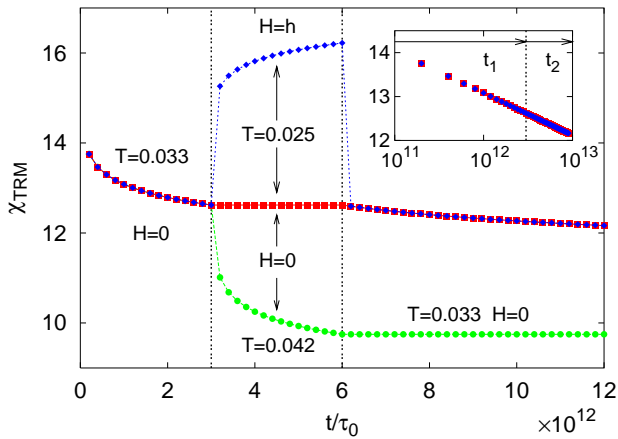}
\caption{(Color online) $\chi_{\rm TRM}$ vs time using the same protocols as in
Figs. 3,4,5 of Sun et al.\cite{Sun03}. The system is cooled to $T=0.033$ 
at the rate of $2.4\cdot 10^{13}$~$\tau_0$ 
per temperature unit in a field which is cut just before
 recording $\chi_{\rm TRM}$. After a time of
$t_1=3\times10^{12}$~$\tau_0$ the temperature is changed. 
The relaxation at the new temperature is recorded either in $H=0$ or $H=h$ 
in period of $t_2=3\times10^{12}$~$\tau_0$. Then the temperature is shifted
back to $T=0.033$ and the field is set to zero. In the inset, $t_1$ and $t_3$ 
parts of $\chi_{\rm TRM}$ with the negative temperature cycling are plotted 
as a function of the total time elapsed at $T=0.033$. 
}
\label{fig3}
\end{figure}

\begin{figure}[b]
\includegraphics[width=\columnwidth]{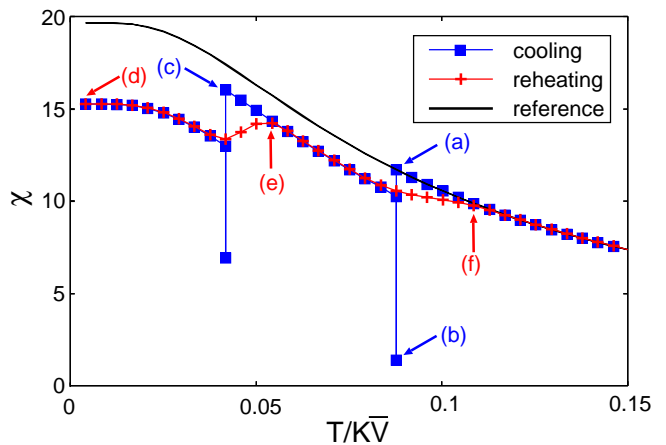}
\caption{(Color online) FC susceptibility vs temperature observed in the same protocol as 
in Fig.~2 of Sun et al.\cite{Sun03}. 
 The field is cut during the intermittent stops 
 of the cooling at $T_1=0.088$ and at $T_2=0.042$ for a period of 
 $10^{14}~\tau_0$. The magnetization in zero field after the waiting time is 
 shown although it was not shown by Sun et al. The arrows in the figure
 indicate at which stages during the procedure we measure and show the
  $M$-spectra in Fig.~\ref{fig5}.
 The cooling (and reheating) rate is the same as that in Fig.~\ref{fig3}.
}
\label{fig4}
\end{figure}

\begin{figure*}[ht]
\includegraphics[width=2\columnwidth,angle=0]{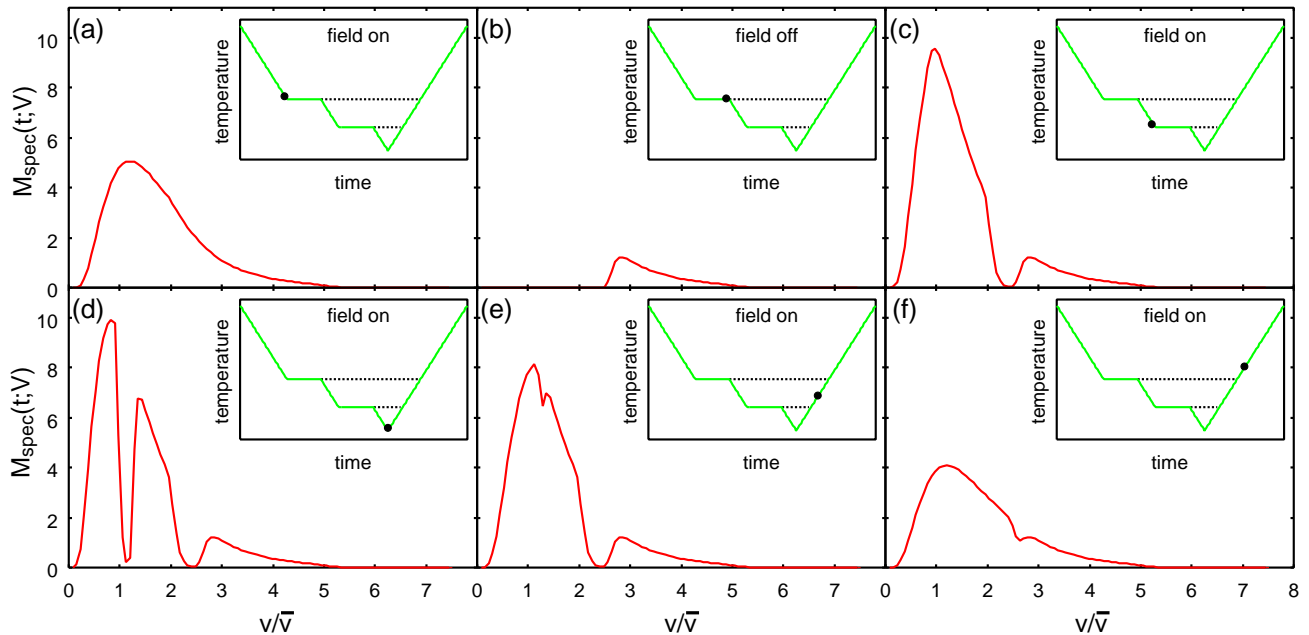}
\caption{(Color online) M-spectra at six representative states which are indicated in 
Fig.~\ref{fig4} by arrows. The point in each inset also shows the time of 
the measurement. }
\label{fig5}
\end{figure*}

As mentioned in subsection~\ref{subsec:2-1}, however, the ZFCM curve is 
independent of $\tw$. One may consider that this $\tw$-independence of 
the ZFCM is a consequence of the simple two-states description of   
our model. Actually, by considering several competing sources of anisotropy 
(for instance magnetocrystalline and magnetostatic energy), 
we can think of a multi-states system with some energy levels 
different from each other. 
Then, the ZFCM of the model should depend weakly on $\tw$ 
even if interactions among particles are absent. 
In fact, we will show in section~\ref{sec:3} that 
the random energy model, which has a huge number of states 
whose energies are different from each other, 
exhibits strong aging in genuine ZFC protocols. 
However, we consider that a significantly small $\tw$-dependence of 
the ZFCM as compared to that of the TRM is one of the characteristic 
properties of superparamagnets since in 
ordinary spin glasses, a strong $\tw$-dependence is observed not only 
in the TRM but also in the ZFCM. 
Therefore, indubitable experimental evidence for spin-glass dynamics in a system can only be found by investigating aging effects in the ZFCM.

From the sum rule for the ZFCM, TRM and FCM, we find 
\begin{equation}
M_{\rm TRM}(t,\tw) = M_{\rm FC}(t+\tw) - M_{\rm ZFC}(t),
\label{eqn:sum-rule}
\end{equation}
where we have used the fact that the ZFCM does not depend on $\tw$ 
in our model. This equation tells us that the $\tw$-dependence of the TRM 
in our model is merely a consequence of slow relaxation of the FCM. 
This is in contrast to ordinary spin glasses where 
the TRM and the ZFCM strongly depend on $\tw$ even if 
$M_{\rm FC}(t')$ for $t' \ge \tw$ hardly relaxes.\cite{MathieuJonsson01}

Another important point is that the peak position of 
the $M$-spectrum in Fig.~\ref{fig2}(a) (and the relaxation rate $S(t)$ 
in Fig.~\ref{fig2}(b)) ceases to shift if 
$\tw \ge \tau_0 \exp(KV_{\rm peak}/\Tm)$, 
where $V_{\rm peak}$ is the peak position of the $M$-spectrum 
in equilibrium (its explicit value is around $1.2$ in the present case).  
On the other hand, aging in spin glasses 
is believed to persist eternally in the thermodynamic limit 
since the relaxation time diverges below the critical temperature.

After the field is cut in the above-mentioned TRM protocol with $\tw =0$, 
we may further introduce some cycling processes,\cite{Sun03,Sahoo02} 
as shown in Fig.~\ref{fig3}. Now let us first consider a negative-temperature 
cycling in zero field. The temperature is changed as 
$\Tm=0.033 \rightarrow T_2=0.025\rightarrow \Tm$. 
Since the blocking volume $V_{\rm B}$ at $T_2$ 
is smaller than that at $\Tm$, the superspins which were dynamically 
active at $\Tm$ are frozen in the second stage at $T_2$, while 
the dynamically active superspins at $T_2$ do not change because 
they were already equilibrated (depolarized) by the first-stage aging 
at $\Tm$. Hence $M_{\rm TRM}$ does not change at all in the second stage
(squares in Fig.~\ref{fig3}). The shape of the $M$-spectrum in 
this stage is essentially the same as that shown in Fig.~\ref{fig5}(b), below. 
After the system comes back to $\Tm$ the relaxation of $M_{\rm TRM}$ 
resumes from the value at the end of the first-stage. If the 
field is applied in the second stage of the above protocol, 
the superparamagnetic and dynamically active 
superspins at $T_2$ respond to it. 
The $M$-spectrum at the end of this stage is essentially the same as that 
in Fig.~\ref{fig5}(c). The induced magnetization in the second stage almost  
immediately disappears in the last stage at $\Tm$ since the superspins which 
carried the excess magnetization are rapidly equilibrated (depolarized) at the 
higher temperature. In the positive-temperature cycling with $T_2=0.042$ 
under $h=0$, superspins which are blocked at $\Tm$ but not at $T_2$ (i.e. 
superspins of nanoparticles whose volume is larger than  
$V_{\rm B}$ at $\Tm$ but smaller than $V_{\rm B}$ at
$T_2$) are rapidly depolarized in the second stage. They are frozen as 
depolarized after changing the temperature back to $\Tm$, and thus  $M_{\rm TRM}$ remains constant at a much smaller value. 
The significant relaxation is expected 
to resume at a time scale when the isothermal  $M_{\rm TRM}$ at $T_{\rm m}$ 
reaches this small value. These features have been in fact observed 
by Sun et al.\cite{Sun03} in a permalloy nanoparticle system.

Lastly let us discuss the peculiar memory effect 
in Fig.~2 of Sun et al.\cite{Sun03}. They introduce 
intermittent stops, at $T_i$, in the FC process and at the same 
time they cut off the field, let the system relax by a certain period 
$t_i$, and then resume the FC process. When the system is reheated  
after reaching a certain low temperature, the magnetization curve 
clearly manifests that the system keeps memories imprinted 
by the preceding FC process. We have applied the 
same protocol to our simple model of superparamagnets, and have 
reproduced qualitatively identical results to theirs as shown 
in Fig.~\ref{fig4}. 

It is clarified in Fig.~\ref{fig5} that this peculiar memory effect originates 
from the blocking of superspins by demonstrating the $M$-spectra of some 
representative instants of the process. After the first stop at $T=T_1$ 
under $h=0$, the $M$-spectrum of Fig.~\ref{fig5}(b) tells us that the blocking 
volume $V_{\rm B1}$ is around $3.0$, namely, the 
superspins of nanoparticles with $V \nle V_{\rm B1}$ are completely 
equilibrated (depolarized), while the frozen superspins of nanoparticles with 
$V \nge V_{\rm B1}$ are still blocked at $T=T_1$ 
after the waiting time. As the FC process is resumed, the memory of the 
first stop at $T=T_1$ is imprinted as a dip at $V \simeq V_{\rm B1}$ in 
the $M$-spectrum [Fig.~\ref{fig5}(c)], since that part of the $M$-spectrum 
is well blocked during the aging at significantly lower temperatures 
than $T_1$. Similarly, by the 
second stop at $T=T_2$ and recooling afterwards, another dip at 
$V_{\rm B2} \simeq 1.3$ is imprinted in the $M$-spectrum as seen in 
Fig.~\ref{fig5}(d). In the reheating process, Fig.~\ref{fig5}(e) and (f) 
illustrate that the frozen part of the $M$-spectrum melts starting from 
small $V$. The consequence is nothing but the memory 
effect reported by Sun et al.

\begin{figure}[b]
\includegraphics[height=\columnwidth,angle=270]{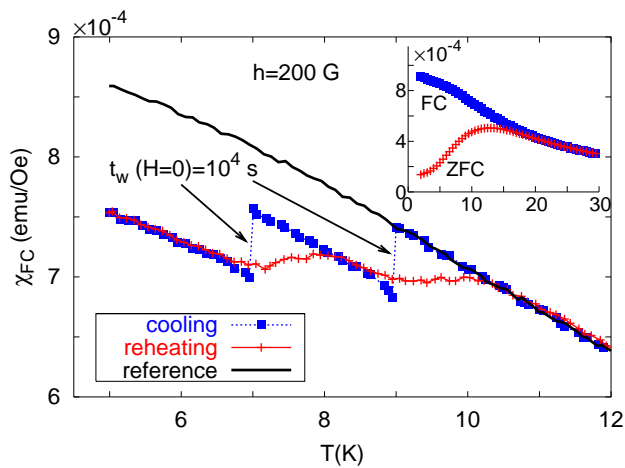}
\caption{(Color online) FC susceptibility of the ferritin with 
the same protocol as that in Fig.~\ref{fig4}. 
The field is cut during the intermittent stops of the cooling at 
$T=9$~K and at $T=7$~K for $10^4$~s at each temperature.
The cooling (and reheating) rate is $1.7\times10^{-3}$~K/s.
The inset shows the ZFC and the FC susceptibilities vs temperature. 
}
\label{Ferritin}
\end{figure}
\begin{figure}[b]
\includegraphics[height=\columnwidth,angle=270]{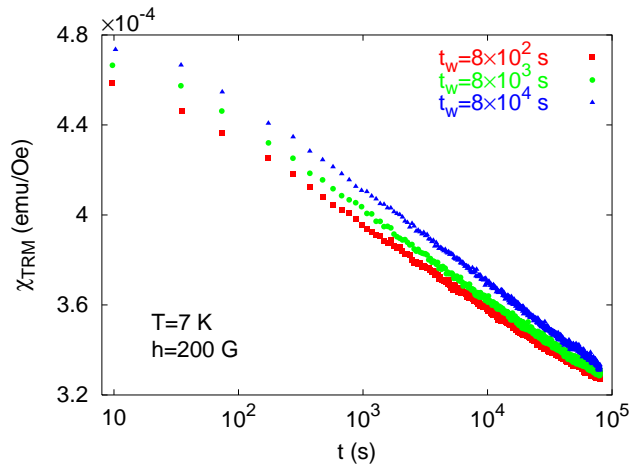}
\caption{(Color online) Relaxation of the TRM susceptibility of the ferritin. 
After the system is cooled to $T=7.0$~K under a $200$~Oe field at
a rate of $0.17$~K/s, it is kept at the temperature 
under the field for $\tw$, and then the field is cut 
and the magnetization decay is measured as a function of the 
elapsed time $t$ after the field is cut. The waiting time $\tw$ is 
$8\times 10^2$, $8\times 10^3$ and $8\times 10^4$ (from left to right). 
}
\label{FerritinTRM}
\end{figure}
\subsection{Experiments on a superparamagnet}
In order to clarify how far our simple model captures the essence of 
{\it real} superparamagnets, we perform experiments on a 
{\it model superparamagnet}, namely natural horse-spleen 
ferritin.\cite{KilcoyneCywinski95,MamiyaNakatani02}
It is an iron-storage protein, and has a spherical cage 8~nm in diameter 
containing polydispersive cores of antiferromagnetic 
ferrihydrite.\cite{Awschalom92,Harrison67} 
Each core has a small magnetic moment of $\sim300$~$\mu_{\rm B}$ due to 
its uncompensated spins.\cite{KilcoyneCywinski95,Makhlouf97} 
%For example, N\'eel-Arrhenius behavior is observed in the ferritin 
%over eight orders of magnitute in frequency.\cite{KilcoyneCywinski95} 
Figure~\ref{Ferritin} shows the result of the memory 
experiment with the same protocol as that in Fig.~\ref{fig4}. 
It is clear from the figure that this superparamagnet also exhibits 
the same memory effect as that observed by Sun et al. 
%as that observed in the permalloy nanoparticle system\cite{Sun03}. 
In fact, this memory behavior is also observed in other 
superparamagnets\cite{ZhengZhang04,ChakravartyFrydman04}. 
The FCM without stops is shown in the inset of Fig.~\ref{Ferritin}. 
We see that the FCM increases monotonically with decreasing temperature. 
As we discussed in the last paragraph of 
subsection~\ref{subsec:2-2}, this is a typical feature 
of superparamagnets. 

Figure~\ref{FerritinTRM} shows relaxation of the TRM susceptibility 
with the same protocol as that in Fig.~\ref{fig2}(B). 
We clearly see that the TRM exhibits a similar $\tw$ dependence 
to that in our simple model of superparamagnets. 
We have also checked a tendency that 
the peak of the relaxation rate $S(t)\equiv -h^{-1}d\log M /d\log t$
shifts to larger times with increasing $\tw$, 
although the data are a bit too noisy to clarify whether the 
peak is located around $\tw$ or not.

We have also done memory experiment in the {\it genuine} 
ZFC protocol.\cite{MathieuJonsson01} 
In this experiment, we measure $\chi_{\rm ZFC}$ which includes 
intermittent stops in the ZFC process and $\chi_{\rm ZFC}^{\rm ref}$ 
without such stops. The stopping temperatures are $9$~K and $7$~K, 
and the period of intermittence is $10^4$~s at each temperature. 
Note that the stopping temperatures are well below the blocking 
temperature $T_{\rm B}\approx 13$~K (see the inset of Fig.~\ref{Ferritin}). 
The cooling (and reheating) rate is the same as that in Fig.~\ref{Ferritin}. 
The result of the experiment is that there is no significant difference 
between $\chi_{\rm ZFC}$ and $\chi_{\rm ZFC}^{\rm ref}$ at any temperatures
(not shown), i.e., no memory is imprinted by the aging 
under zero field. This is also the expected result for superparamagnets.

\section{Slow Dynamics in Superspin Glasses}
\label{sec:3}
\begin{figure}[thb]
\includegraphics[width=\columnwidth]{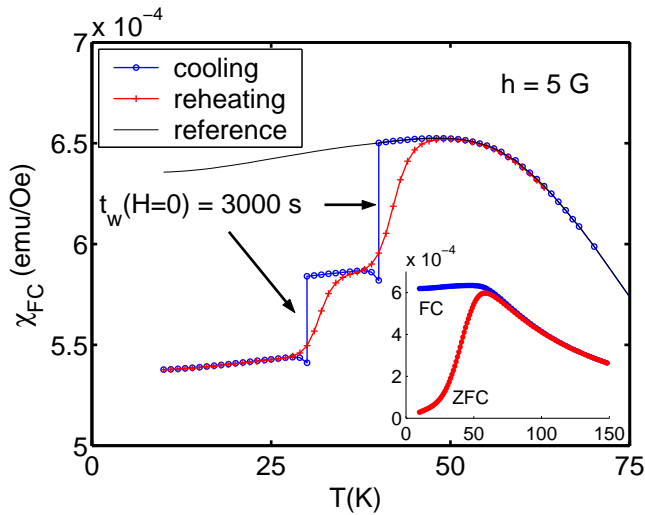}
\caption{(Color online) FC susceptibility of the Fe-N system with 
the same protocol as that in Fig.~\ref{fig4}. 
%~\cite{sunetal2003}. 
The critical temperature of the sample is around 60~K.\cite{MamiyaNakatani99B} 
The field is cut during the intermittent stops of the cooling at 
$T=40$~K and at $T=30$~K for 3000~s at each temperature.
The cooling (and reheating) rate is $0.01$~K/s.
The inset shows the ZFC and the FC susceptibilities vs temperature. 
\label{FENfcmemory}}
\end{figure}

Various memory experiments are performed on a dense 
Fe-N ferromagnetic nanoparticle system which has been shown to be a superspin 
glass.\cite{MamiyaNakatani98,MamiyaNakatani99,MamiyaNakatani99B,JonssonYoshino04} 
Figure~\ref{FENfcmemory} shows the result of the memory experiment following 
the protocol as that in Fig.~\ref{fig4}. At the intermittent stops of the FC 
process, while the field is set to zero, the value of the magnetization 
decreases. On the subsequent reheating, the magnetization value in the 
preceding cooling process is recovered, for each stop, at a temperature a 
bit above that of the stop. At a glance, the memory effect in this superspin 
glass is qualitatively the same as that in superparamagnets 
indicating a similar origin of the effect. Another interesting observation in 
Fig.~\ref{FENfcmemory} is that the FCM of the Fe-N system 
after resuming the cooling behaves almost in 
parallel to the FCM without the intermittent stops (reference curve) 
though its absolute magnitude is significantly smaller than the latter. This 
feature is also seen for the superparamagnets as shown in Figs.~\ref{fig4} and 
\ref{Ferritin}, and so it suggests that the mechanism behind 
the memory effect is also common. 
%%, i.e., blocking of `moments' with relatively large (free) energy barriers. 

\begin{figure}[thb]
\includegraphics[width=\columnwidth]{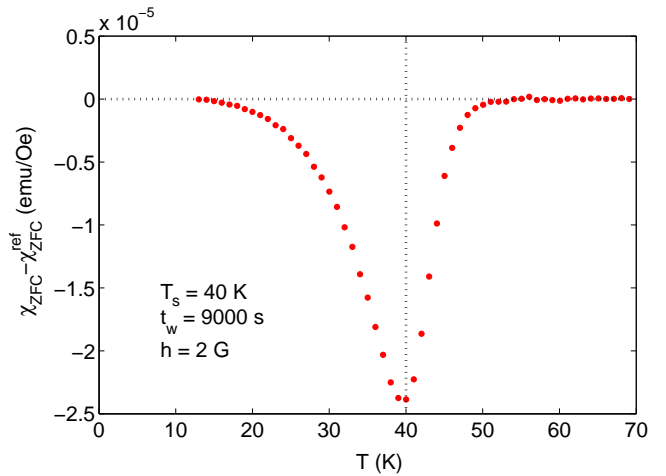}
\caption{(Color online) Difference of the ZFC susceptibility of the the Fe-N system. 
The ZFC process is intermitted at $T=40$~K for 9000~s 
in the measurement of $\chi_{\rm ZFC}$, while $\chi_{\rm ZFC}^{\rm ref}$ 
is measured without such a stop. The cooling rate is $0.1$~K/s, 
and the reheating rate is $0.01$~K/s.  
\label{zfcmemory}}
\end{figure}

Now let us go into further comparisons between the results so far obtained
for the superparamagnets and those for the superspin glass. One significant 
difference between the two is seen in the behavior of the reference FCM 
without the intermittent stops. The FCM of the Fe-N system does not 
increase but even decreases as the temperature is decreased. According to 
the argument in the last paragraph of subsection~\ref{subsec:2-2}, 
this implies that the Fe-N system is in fact not a 
superparamagnet also in this respect. Actually the nearly constant FCM is 
considered to be a typical property of ordinary spin glasses. 
A further important phenomenon which is peculiar to superspin glasses 
is memory effect in the {\it genuine} ZFC protocol. 
Figure~\ref{zfcmemory} 
shows an experimental result of the Fe-N system where 
the difference between the ZFCM's with and without an intermittent stop 
at $T_{\rm s}$ in the cooling process is presented. The difference is 
clearly observed as a dip at $T \simeq T_{\rm s}$. 

\begin{figure}[t]
\includegraphics[height=\columnwidth,angle=270]{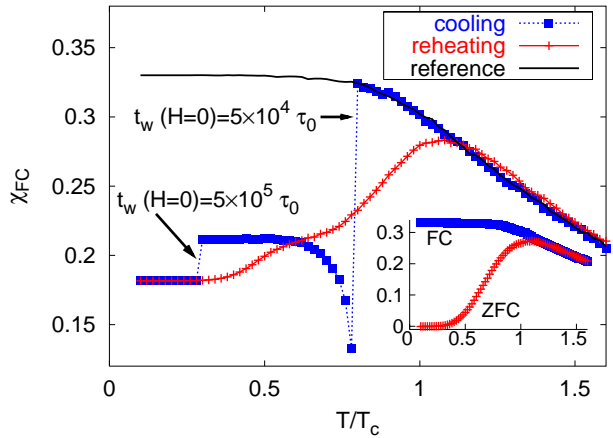}
\caption{(Color online) FC susceptibility of the REM with 
the same protocol as that in Fig.~\ref{fig4}. 
The field is cut during the intermittent stops of the cooling 
at $T=0.8~T_{\rm c}$ for $5\times10^4~\tau_0$ and 
at $T=0.3~T_{\rm c}$ for $5\times10^5~\tau_0$, where $\tau_0$ is a 
microscopic time of the model. The cooling (and reheating) rate is 
$2\times10^{-5}~T_{\rm c}/\tau_0$. The inset shows the ZFC 
and the FC susceptibilities vs temperature. 
\label{REMfcmemory}}
\end{figure}

\begin{figure}[t]
\includegraphics[height=\columnwidth,angle=270]{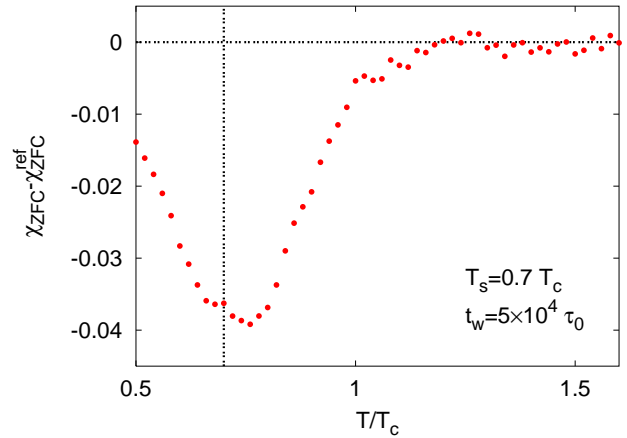}
\caption{(Color online) Difference of the ZFC susceptibility of the the REM. 
The ZFC process is intermitted at $T=0.7~T_{\rm c}$ for $5\times10^4~\tau_0$ 
in the measurement of $\chi_{\rm ZFC}$, while $\chi_{\rm ZFC}^{\rm ref}$ 
is measured without such a stop. The cooling (and reheating) rate is 
the same as that in Fig.~\ref{REMfcmemory}. 
\label{REMzfcmemory}}
\end{figure}

Now let us discuss possible theoretical interpretations of 
these experimental results. The first theoretical model we consider is 
the random energy model (REM).\cite{Derrida81,Bouchaud92,BouchaudDean95} 
The REM consists of a huge number of states. The barrier energy $E_{\rm B}$, 
which the system needs to overcome in order to go to a new state, 
is assigned to each state randomly and independently according to the 
exponential distribution 
$\rho(E_{\rm B})=1/T_{\rm c} \exp[-E_{\rm B}/T_{\rm c}]$. Since the 
average relaxation time $\langle \tau \rangle=\int_0^{\infty}{\rm d}E_{\rm B}
\rho(E_{\rm B})\tau_0 \exp(E_{\rm B}/T)$ diverges below $T_{\rm c}$, the REM shows various memory and 
aging behavior in the low temperature phase.\cite{Bouchaud92,BouchaudDean95,SasakiNemoto00c} 
Let us now see to what extent the experimental results shown in this section 
are reproducible by the REM. First, Fig.~\ref{REMfcmemory} shows 
the result with the same protocol as that in Fig.~\ref{fig4}. 
The result is qualitatively rather similar to that of the Fe-N system 
shown in Fig.~\ref{FENfcmemory}. 
In particular, it should be emphasized that 
the flatness of the FCM below the critical temperature, which can not be 
captured by our simple model of superparamagnets, 
is reproduced in the REM. Second,  Fig.~\ref{REMzfcmemory} shows the result 
of simulation which corresponds to the ZFC memory experiment 
in Fig.~\ref{zfcmemory}. 
Again, the result is qualitatively very similar to that in the experiment. 
A crucial property of the REM to understand this result is that the system 
goes into deeper and deeper states with higher and higher energy barriers as time 
progresses.\cite{Bouchaud92,SasakiNemoto00b} Therefore, the typical 
energy barrier of the state in which the system is blocked depends on how long 
the system has been aged at a low temperature. 
Since it is more difficult for the system blocked in a state 
with a higher energy barrier to respond to the field, 
the difference of the typical energy barrier of the 
state in which the system  is blocked with and without intermittent stop on cooling causes the dip in Fig.~\ref{REMzfcmemory}.

We have seen that the experimental results are well reproduced by the REM. 
However, the link between each state in the REM and an actual spin
configuration in the system is not so clear. On the other hand, the
droplet theory\cite{FisherHuse88B,FisherHuse88} gives some insight into
spin configurations in the (nonequilibrium) dynamics of real spin
glasses. For example, after a spin glass is rapidly quenched in a
field $h$ to a temperature $T$ below $\Tc$, spin-glass domains, or
clusters, which are
in local equilibrium with respect to $(T, h)$ are considered to grow. At
a certain instance $t$ after the quench, clusters with various volumes $\Vc$
or linear sizes $L\ (\sim \Vc^{1/d})$ exist. We may think of their
distribution $P(t;\Vc)$, analogously to $P(V)$ in the previous section.
Furthermore, in the droplet theory, each cluster of a size $L$ is
considered to flip by a thermal activation process whose mean energy barrier 
$B_L$ is a function of $L$ ($B_L \sim L^\psi$ in the original
droplet theory\cite{FisherHuse88}). The thermally activated process governs 
the response of clusters to an applied field. This 
situation is rather similar to the
two-state description of the superparamagnet, and we may expect that the
magnetization of the spin glass is also described by 
Eq.(\ref{eqn:ave-mag}), though the functional form of $M(t;V)$ has to be 
properly modified and $P(V)$ has to be replaced by a 
time-dependent distribution $P(t;\Vc)$. We also note that the above
argument on an atomic spin glass can be directly applied to a superspin
glass if an atomic spin in the former is replaced by a superspin in the
latter.  

An interesting prediction of the droplet theory is the instability of
the equilibrium spin-glass phase under a static magnetic field $h$ of
any strength. This is one of the fundamental issues which has been
debated since the early stage of the spin-glass study and has not been
settled yet. Quite recently, in numerical analysis 
of the field-shift aging protocol, one of the present authors (HT) and
Hukushima~\cite{TakayamaHukushima03} have found results which strongly 
support the prediction of the droplet theory. Here let us argue about our 
experimental results on the superspin glass from this point of view, namely, 
the FCM measured at $T < \Tc$  is not an equilibrium property under $h$ but 
due to the blocking of superspin clusters introduced above.

As noted before, the FCM of a superparamagnet increases with decreasing
$T$. That of the present superspin glass, on the other hand, is nearly
constant at $T \nle \Tc$ as seen in the inset of Fig.~\ref{FENfcmemory}. 
The latter is naturally attributed to the expected fact that the free
energy difference between the two states of a superspin cluster is given
not only by the Zeeman energy but also by the residual interactions
between the cluster and its surroundings (the stiffness energy of a cluster
in the droplet theory). If the field strength is sufficiently small,
which is the case of the present interest, the latter certainly
dominates the Zeeman energy. Therefore, when the cluster is blocked, its
magnetization points either in parallel or antiparallel to the field 
direction. Consequently the branch of the FCM at $T \nle \Tc$ in 
Fig.~\ref{FENfcmemory} becomes nearly constant when the temperature is
decreased. 

By further inspection of Fig.~\ref{fig1} and the inset of 
Fig.~\ref{FENfcmemory} we notice that the FCM of a superparamagnet
changes rather smoothly around the blocking temperature, while that of the
superspin glass exhibits a kink-like shape at $T \sim \Tc$. The latter
can be attributed to the time development of $P(t;\Vc)$ which is absent in
a superparamagnet. In fact, in the droplet theory, the rates of growth of
the spin-glass clusters and so of their barrier energy are expected to
be most sensitive to a small change in temperature at $T \simeq \Tc$,
since they are governed by the critical dynamics associated with the
spin-glass transition at $T=\Tc$ under $h=0$. Consequently, even a small
temperature decrease at this temperature range gives rise to an
apparently sharper blocking of superspin clusters. At significantly
lower temperatures than $\Tc$, the thermal activation process governs
dynamics of superspins and yields an almost constant FCM as
described just above. 

Let us turn to memory effects in the genuine ZFC protocol, which are not 
observed in superparamagnets. As we mentioned above, at $T \nle \Tc$,
sizes of clusters are growing as time elapses which gives rise to a
history dependence of $P(t;\Vc)$ in the language of our two-states
model. Since the change of $P(t;\Vc)$ proceeds even in a vanishing field,
memory effects are observed even in the genuine ZFC protocol. 

Lastly one comment is in order on possible differences in slow dynamics
of superspin glasses and atomic spin glasses. As mentioned above,
qualitative aspects of the two are considered to be almost common to each
other. Quantitatively, however, the unit time of a superspin flip 
depend on $T$ and is much larger than the temperature independent atomic-spin flip time. This difference often causes apparent qualitative
differences in the non-equilibrium phenomena in the two spin glasses
because of the common experimental time window, $10^1 \sim 10^5$ s, 
which may differ very much when measured in the unit 
time of each system.\cite{JonssonYoshino04} 

\section{Summary}
\label{sec:4}
We have studied dynamics of superparamagnets by investigating 
a simplified two-states model analytically and ferritin experimentally. 
As a consequence, we have found that
\begin{itemize}
\item[a)] The TRM exhibits a $\tw$-dependence, and the logarithmic 
time derivative of $\chi_{\rm TRM}(t,\tw)$ has a peak around $t\approx\tw$, as observed 
in spin glasses. 
\item[b)] All the experimental results reported by Sun et al.\cite{Sun03} 
are qualitatively reproducible. 
\end{itemize}
In superparamagnets, these aging and memory effects originate 
solely from a broad distribution of relaxation times 
which comes from that of the anisotropy energy barriers. 
The mechanism of these results is well understood by investigating the time 
dependence of the $M$-spectrum (the integrand in Eq.(\ref{eqn:ave-mag})). 
Thus the aging and memory effects a-b) are not a sufficient
proof for the existence of spin-glass dynamics.

We have also studied aging and memory effects in a dense 
Fe-N  nanoparticle system (a {\it superspin glass}) 
experimentally. By comparing the results with those for superparamagnets, 
the following differences have been found:
\begin{itemize}
\item[1)] The FCM of the Fe-N system does not increase but even decreases 
as the temperature is decreased, while the FCM of superparamagnets 
always increases with decreasing temperature. 
\item[2)] In the Fe-N system, the genuine ZFCM also depends on the waiting 
time. Such a $\tw$-dependence in the ZFCM is hardly expected in 
superparamagnets.
\end{itemize}
From the viewpoint of 1), we consider that the permalloy nanoparticle system 
studied by Sun et al. is closer to a superparamagnet, while the Fe-N system 
studied in the present work and the Co-Fe nanoparticle system studied by Sahoo 
et al.\cite{Sahoo02b,Sahoo02,Sahoo03} are closer 
to a superspin glass. 
Lastly, we have argued that these two aspects peculiar 
to superspin glasses are qualitatively reproduced by the random energy model, 
and are well interpreted by the droplet theory in the field of 
the spin-glass study. 

In conclusion, similarities as well as crucial differences in aging and
memory effects in superparamagnets and superspin glasses have been
clarified. In order to distinguish the two types of slow dynamics we
have to choose appropriate aging protocols such as a ZFC process with
intermittent stops of the cooling properly scheduled.

\acknowledgments
M.S. and P.E.J. acknowledge financial support from the Japan Society for the 
Promotion of Science. The present work is supported by a Grant-in-Aid for 
Scientific Research Program (\# 14540351) and NAREGI Nanoscience Project 
from the Ministry of Education, Culture, Sports, Science and Technology. 

\bibliographystyle{prsty}
\bibliography{references2}

\addcontentsline{toc}{chapter}{\protect\bibname}
\begin{thebibliography}{10}

\bibitem{Struik78}
L.~C.~E. Struik, {\em Physical aging in amorphous polymers and other
  materials.} (Elsevier, Houston, 1978).

\bibitem{Lundgren83}
L. Lundgren, P. Svedlindh, P. Nordblad, and O. Beckman, Phys. Rev Lett. {\bf
  51},  911  (1983).

\bibitem{BellonCiliberto99}
L. Bellon, S. Ciliberto, and C. Laroche,   (1999), cond-mat/9905160.

\bibitem{BellonCiliberto00}
L. Bellon, S. Ciliberto, and C. Laroche, Europhys. Lett. {\bf 51},  551
  (2000).

\bibitem{RosselMaeno89}
C. Rossel, Y. Maeno, and I. Morgenstern, Phys. Rev. Lett. {\bf 62},  681
  (1989).

\bibitem{Josserand00}
C. Josserand, A.~V. Tkachenko, D.~M. Mueth, and H.~M. Jaeger, Phys. Rev. Lett.
  {\bf 85},  3632  (2000).

\bibitem{VincentBouchaud95}
E. Vincent, J.~P. Bouchaud, J. Hammann, and F. Lefloch, Phil. Mag. B {\bf 71},
  489  (1995).

\bibitem{NordbladSvedlindh98}
P. Nordblad and P. Svedlindh,  in {\em Spin Glasses and Random Fields}, edited
  by A.~P. Young (World Scientific, Singapore, 1998).

\bibitem{JonasonVincent98}
K. Jonason {\it et~al.}, Phys. Rev. Lett. {\bf 81},  3243  (1998).

\bibitem{CugliandoloKurchan93}
L.~F. Cugliandolo and J. Kurchan, Phys. Rev. Lett. {\bf 71},  173  (1993).

\bibitem{BouchaudCugliandolo98}
J.-P. Bouchaud, L. Cugliandolo, J. Kurchan, and M. M{\'e}zard,  in {\em Spin
  Glasses and Random Fields}, edited by A.~P. Young (World Scientific,
  Singapore, 1998).

\bibitem{YoshinoLemaitre00}
H. Yoshino, A. Lema\^{\i}tre, and J.-P. Bouchaud, Eur. Phys. J. B. {\bf 20},
  367  (2000).

\bibitem{SasakiNemoto00}
M. Sasaki and K. Nemoto, J. Phys. Soc. Jpn {\bf 69},  2283  (2000).

\bibitem{KomoriYoshino00}
T. Komori, H. Yoshino, and H. Takayama, J. Phys. Soc. Jpn. Suppl A {\bf 69},
  228  (2000).

\bibitem{BerthierBouchaud02}
L. Berthier and J.-P. Bouchaud, Phys. Rev. B {\bf 66},  054404  (2002).

\bibitem{MezardParisi84}
M. M{\'e}zard {\it et~al.}, J. Physique {\bf 45},  843  (1984).

\bibitem{MezardParisi84b}
M. M\'ezard {\it et~al.}, Phys. Rev. Lett. {\bf 52},  1156  (1984).

\bibitem{FisherHuse86}
D.~S. Fisher and D.~A. Huse, Phys. Rev. Lett. {\bf 56},  1601  (1986).

\bibitem{BrayMoore87}
A.~J. Bray and M.~A. Moore, Phys. Rev. Lett. {\bf 58},  57  (1987).

\bibitem{FisherHuse88}
D.~S. Fisher and D.~A. Huse, Phys. Rev. B {\bf 38},  386  (1988).

\bibitem{WillsDupuis00}
A. Wills {\it et~al.}, Phys. Rev. B {\bf 62},  R9264  (2000).

\bibitem{DupuisVincent02b}
V. Dupuis {\it et~al.}, J. Appl. Phys. {\bf 91},  8384  (2002).

\bibitem{MathieuSvedlindh00}
R. Mathieu, P. Svedlindh, and P. Nordblad, Europhys. Lett. {\bf 52 (4)},  441
  (2000).

\bibitem{AlberichDoussineau97}
F. Alberich, P. Doussineau, and A. Levelut, J. Phys. I (France) {\bf 7},  329
  (1997).

\bibitem{AlberichBouchaud98}
F. Alberici-Kious {\it et~al.}, Phys. Rev. Lett. {\bf 81},  4987  (1998).

\bibitem{LehenyNagel98}
R.~L. Leheny and S.~R. Nagel, Phys. Rev. B {\bf 57},  5154  (1998).

\bibitem{JonssonMattsson95}
T. Jonsson {\it et~al.}, Phys. Rev. Lett. {\bf 75},  4138  (1995).

\bibitem{MamiyaNakatani99}
H. Mamiya, I. Nakatani, and T. Furubayashi, Phys. Rev. Lett. {\bf 82},  4332
  (1999).

\bibitem{JonssonHansen00}
P. J{\"o}nsson, M.~F. Hansen, and P. Nordblad, Phys. Rev. B {\bf 61},  1261
  (2000).

\bibitem{Sahoo02}
S. Sahoo {\it et~al.}, J. Phys. C {\bf 14},  6729  (2002).

\bibitem{Sahoo03}
S. Sahoo {\it et~al.}, Phys. Rev. B {\bf 67},  214422  (2003).

\bibitem{Jonsson04}
P.~E. J{\"o}nsson, Adv. Chem. Phys. {\bf 128},  191  (2004).

\bibitem{BatlleLabarta2002}
X. Batlle and A. Labarta, J. Phys. D: Appl. Phys. {\bf 35},  R15  (2002).

\bibitem{LuoNagel91}
W. Luo, S.~R. Nagel, T.~F. Rosenbaum, and R.~E. Rosensweig, Phys. Rev. Lett.
  {\bf 67},  2721  (1991).

\bibitem{JonssonSvedlindh98}
T. Jonsson, P. Svedlindh, and M.~F. Hansen, Phys. Rev. Lett. {\bf 81},  3976
  (1998).

\bibitem{MamiyaNakatani99B}
H. Mamiya and I. Nakatani, Nanostruct. Mater. {\bf 12},  859  (1999).

\bibitem{Sahoo02b}
S. Sahoo {\it et~al.}, Phys. Rev. B {\bf 65},  134406  (2002).

\bibitem{Sun03}
Y. Sun, M.~B. Salamon, K. Garnier, and R.~S. Averback, Phys. Rev. Lett. {\bf
  91},  167206  (2003).

\bibitem{SasakiJonsson03}
M. Sasaki, P.~E. J{\"o}nsson, H. Takayama, and P. Nordblad,   (2003),
  cond-mat/0311264 (to be published).

\bibitem{ZhengZhang04}
R.~K. Zheng and X.~X. Zhang,   (2004), cond-mat/0403368.

\bibitem{ChakravartyFrydman04}
S. Chakravarty {\it et~al.},   (2004), cond-mat/0403574.

\bibitem{KilcoyneCywinski95}
S.~H. Kilcoyne and R. Cywinski, J. Magn. Magn. Mater. {\bf 140-144},  1466
  (1995).

\bibitem{MamiyaNakatani02}
H. Mamiya, I. Nakatani, and T. Furubayashi, Phys. Rev. Lett. {\bf 88},  067202
  (2002).

\bibitem{MamiyaNakatani98}
H. Mamiya, I. Nakatani, and T. Furubayashi, Phys. Rev. Lett. {\bf 80},  177  (1998).

\bibitem{JonssonYoshino04}
P.~E. J{\"o}nsson, H. Yoshino, H. Mamiya, and H. Takayama,   (2004),
  cond-mat/0405276.

\bibitem{Parisi83}
G. Parisi, Phys. Rev. Lett. {\bf 50},  1946  (1983), and references therein.

\bibitem{FisherHuse88B}
D.~S. Fisher and D.~A. Huse, Phys. Rev. B {\bf 38},  373  (1988).

\bibitem{TakayamaHukushima03}
H. Takayama and K. Hukushima,   (2003), cond-mat/0307641.

\bibitem{Derrida81}
B. Derrida, Phys. Rev. B {\bf 24},  2613  (1981).

\bibitem{Bouchaud92}
J.~P. Bouchaud, J. Phys. I France {\bf 2},  1705  (1992).

\bibitem{BouchaudDean95}
J.-P. Bouchaud and D. Dean, J. Phys. I, France {\bf 5},  265  (1995).

\bibitem{KlikChang94}
I. Klik, C.-R. Chang, and J. Lee, J. Appl. Phys. {\bf 75},  5487  (1994).

\bibitem{MathieuJonsson01}
R. Mathieu, P. J{\"o}nsson, D.~N.~H. Nam, and P. Nordblad, Phys. Rev. B {\bf
  63},  092401  (2001).

\bibitem{Awschalom92}
D.~D. Awschalom {\it et~al.}, Phys. Rev. Lett. {\bf 68},  3092  (1992).

\bibitem{Harrison67}
P.~M. Harrison, F.~A. Fischbach, T.~G. Hoy, and G.~H. Haggis, Nature (London)
  {\bf 216},  1188  (1967).

\bibitem{Makhlouf97}
S.~A. Makhlouf, F.~T. Parker, and A.~E. Berkowitz, Phys. Rev. B {\bf 55},
  R14717  (1997).

\bibitem{SasakiNemoto00c}
M. Sasaki and K. Nemoto, J. Phys. Soc. Jpn {\bf 69},  3045  (2000).

\bibitem{SasakiNemoto00b}
M. Sasaki and K. Nemoto, J. Phys. Soc. Jpn {\bf 69},  2642  (2000).

\end{thebibliography}

\end{document}